# Oncoprotein Metastasis Disjoined

**Razvan Tudor Radulescu**

*Molecular Concepts Research (MCR), Munich, Germany*
E-mail: ratura@gmx.net

ABSTRACT

As the past decade barely dawned, a fundamentally novel view of cancer relating to signal transduction through intracellular hormones/growth factors and their subunits began to unfold. Further along, it gained additional substance with the advent of the interdisciplinary fields of particle biology and peptide strings which explain (onco)protein dynamics in spacetime, for instance insulin-driven sub- and trans-cellular carcinogenesis, by physical principles. Here, this new understanding is expanded to introduce the concept of "oncoprotein metastasis" preceding cancer cell spread and, thereby, a particular emphasis is placed on its potential role in the emergence of the pre-metastatic niche. Consistent with this perception, yet unlike currently advocated treatments that target cancer cells only, future antineoplastic strategies should aim to mimic natural tumor suppressors as well as involve both (morphologically) normal and malignant cells. If validated in human patients with advanced cancer disease, its otherwise frequently lethal course may be halted and reversed just in time.

A huge amount of experimental data relating to the problem of cancer metastasis has accumulated (1). By contrast, concepts providing a basic interpretation of these observations have been few. Here, a new scenario is put forward on the spreading of the neoplastic process across cells and tissues that may prove seminal both for our future understanding and treatment of malignancies.

Currently, three major views serve to explain both the emergence of cancer and its metastatic spread. One of these frameworks is the (cellular) oncogene theory developed by Michael Bishop and Harold Varmus along with Dominique Stehelin in the early 1970s and still prevailing to date (2). It assumes an overactivity of oncogenes (through gene amplification and/or mutation) being at the root of neoplastic transformation and cancer cell spread.

Another mechanism proposed to initiate and expand primary tumors as well as trigger its associated metastasis is the loss of (individual) tumor suppressor activity, a proposal that, in its essence, can be traced back to Theodor Boveri as well as later on to Alfred Knudson´s two-hit hypothesis and is best exemplified by the first discovered tumor suppressor gene *Rb* (3,4). This tumor suppressor dysfunction that is thought to decisively contribute to carcinogenesis can occur at the level of the genes, but, at least as far as the key tumor suppressor retinoblastoma protein (RB) is concerned, is found even more frequently at the protein level in the form of post-translational modifications (5).

Thirdly and in more recent years, chromosomal abnormalities resulting in aneuploidy have emerged as another important potential origin for cancer (metastasis), mainly through the work of Peter Duesberg and his associates (6).

Yet, genetic and chromosomal views of cancer are not entirely satisfactory. For instance, Richmond Prehn raised already in 1994 the question as to whether mutations found in some genes of cancer cells may be rather an epiphenomenon accompanying the (already initiated) neoplastic process than a causative event (7). Along the same lines, Judah Folkman and coworkers have also appealed to (additionally) go outside the cancer genome (8) to find broader answers, one direction being traced by his long-standing idea on tumor angiogenesis (9). Nevertheless, such example may be understood just as an incentive to embark upon a multidirectional conceptual journey beyond the genome that altogether should yield the ultimate solution to the cancer problem.

In this context, my peptide string theory (10-12) is likely to represent a significant addition. It rests upon the assumption according to which major biological processes concerning distinct, yet related proteins are the result both of (long-distance) attractive forces in the sense of the physical string theory (13,14) and of "emergent properties" inherent to the same proteins whereby the term

"emergent" is to be understood as employed by John Searle in his book entitled "The Mystery of Consciousness" where he illustrated it by explaining that single $H_2O$ molecules are not fluid by themselves, yet they create water when joining each other which is thus an "emergent" property of theirs. More specifically speaking, I have outlined how, for instance, the oncogenic information or message, respectively, encoded by certain peptide motifs in oncoproteins "propagates" within and across cells as a result of numerous interactions among (allosteric) proteins (silently or dormantly) harboring similar or identical motifs, hence yielding an "oncogenic state" in (normal as well as transformed) tissues and ultimately in an entire organism affected by a given cancer disease (11,12).

From such (onco)peptide string view of cancer, it follows that, contrary to the current dogma of solely cancer cell-targeted therapies (15), both normal and cancer cells should be treated, specifically aiming to globally disrupt the oncoprotein-driven **process** that is not limited to individual (cancer) cell borders (16,17). This could be achieved by cell-permeable tumor suppressor peptides, particularly those using tumor-suppressive components of the RB pathway as a template, that enter **both** normal and cancer cells, as already suggested (18).

At this point, it may be worthwhile considering the conceptual origins of my peptide string theory and its (anticipated) equivalents in the work of other investigators. Following my initial identification of the LXCXE RB-binding motif in the B-chain of human insulin in 1992 (19), I predicted a physical interaction between insulin and RB as well as the involvement of this complex formation in embryogenesis and oncogenesis (19). I then generalized this potential phenomenon to postulate that the increased (post-translational) generation of hormone and growth factor subunits may contribute to (accelerated) tumor formation (20), thus bypassing lengthy cell membrane-to-nucleus signal transduction cascades and ensuing activations of various gene expression machineries. Furthermore, I surmised that there may exist also other examples besides insulin for such "nucleocrine" hormones and growth factors which, as I defined them by this term, directly bind nuclear tumor suppressors, thereby inactivate them and hence promote neoplastic transformation (21).

In the course of this early work on the nuclear insulin-RB interaction as well as, more broadly speaking, on growth factors and/or their subunits "gravitating" towards the (cancer) cell nucleus, I realized in 1994 that some physical laws may particularly apply to carcinogenesis and summarized such thoughts in my particle biology theory published in 2003 (22). This framework and particularly its intrinsic "bio-gravitation" and field concepts then served as a basis for the development of my peptide string theory which could be considered as a specific candidate for the different law in physics postulated to exist in biology by Max Delbrück a long time ago.

A key twist of this retrospective on the roots of my peptide string concept is the fact that I had already envisaged within my particle biology theory in 1994 that oncogenic **mechanisms**- and subsequently also equivalent (epi)genetic changes- occur in morphologically normal cells during a given cancer disease as well as inferred (protein-based) **fields** transcending cellular boundaries (22). Consistent with this anticipation, Helene S. Smith *et al.* reported in 1996 that normal cells adjacent to a breast carcinoma bear neoplastic changes such as a loss of heterozygosity in some candidate tumor suppressor genes and thereby conjectured a potential "field effect" (23). Likewise, it was revealed by infrared spectroscopy that morphologically normal cells contiguous to neoplastic tissues display structural abnormalities (24) or, respectively, feature a cancer DNA phenotype (25).

Consequently, if normal cells in a cancer patient contain such pre-malignant alterations, it would not be too far-fetched to assume that they also harbor (as yet undetermined or poorly determined) single protein modifications (such as a hyper-phosphorylation and thus inactivation of RB) and oncoprotein-tumor suppressor complexes that may predispose these cells to hyperproliferation and gradually also neoplastic transformation.

In this context, the elevated (fasting) insulin blood levels or hyperinsulinemia, respectively, observed in some cancer patients (26-31) could be an important clinical hint to check for the above presumed protein-based phenomena in normal cells adjacent to a given carcinoma. More precisely stated, such abundance of circulating insulin may serve as one source among several (besides potential paracrine and autocrine ones) to fuel insulin for cellular internalization and nuclear translocation in many different tissues and, in turn, nuclear insulin could then bind and inactivate RB, as already shown (32). Considering further that this growth-promoting hormone is able to undergo a rapid transport across cells (33), those insulin molecules remained unbound to nuclear RB during a first oncoprotein invasion "wave" could proceed to enter and, due to their still excessive concentrations, transform other normal cells.

These events could occur before any (epi)genetic and/or morphological changes have taken place in these cells, hence being even more in advance of the seed-and-soil process Stephen Paget had hypothesized already in the 19th century to involve individual cancer cells leaving the primary tumor for the bloodstream and, upon extravasation, settling down in distant tissue sites conducive to their thriving. Such proposed oncoprotein-driven mechanism would crucially contribute to establishing the pre-metastatic niche whose formation is a prerequisite for the invasion and subsequent survival of metastatic cells arriving from far away. Conceivably, these "molecular preparations" could, however, only be completed in those tissues that are not capable of successfully mounting tumor-suppressive mechanisms, hence providing a novel dimension to explaining the clinically observed phenomenon according to which some primary tumors preferentially and visibly metastasize to certain sites in the body. Moreover, those cancer cells which have achieved to "land" in distant tissues should accelerate the phenotypic conversion of their surrounding pre-malignant cells into overt tumor cells, e.g. through soluble factors they may secrete into their microenvironment.

Essentially, the hyperinsulinemia of malignancy and its intracellular sequelae would altogether represent the paradigm for what I wish to coin as "oncoprotein metastasis" which, by my present definition, precedes cancer cell metastasis. In other words, the message carried by the insulin LXCXE motif and its spread across cells in the entire organism would epitomize a paramount oncogenic peptide string.

If this scenario proved true, then the specific disruption of (intracellular) insulin-RB heterodimers through cell-permeable RB-derived peptides (34-36)- that, by virtue of their structure, are not restricted to recognizing only malignant cells- could be one way for effectively treating cancer metastasis. Ultimately, this type of pharmacological agents could serve as a model for future therapies that do no longer target tumor cells alone to eradicate life-threatening metastasis, but instead anticipate and thus most likely disconnect it in time by installing a tumor-suppressive (anti-oncogenic peptide string) state in all cells, normal and cancerous alike.